# Influence of dislocations on the spatial variation of microstructure in martensites

R. Gröger[1,a] and T. Lookman[2,b]

[1]Institute of Physics of Materials, Academy of Sciences of the Czech Republic,
Žižkova 22, 616 62 Brno, Czech Republic

[2]Theoretical Division, Los Alamos National Laboratory,
MS B262, Los Alamos, NM 87545, USA

[a]groger@ipm.cz, [b]txl@lanl.gov

**Abstract.** The continuum theory of dislocations, as developed predominantly by Kröner and Kosevich, views each dislocation as a source of incompatibility of strains. We show that this concept can be employed efficiently in the Landau free energy functional to develop a mean-field mesoscopic model of materials with dislocations. The order parameters that represent the distortion of the parent phase (often of cubic symmetry) are written in terms of elastic strains which are themselves coupled by incompatibility constraints. Since the "strength" of the incompatibility depends on the local density of dislocations, the presence of dislocations affects the evolution of the microstructure and vice versa. An advantage of this formulation is that long range anisotropic interactions between dislocations appear naturally in the formulation of the free energy. Owing to the distortion of the crystal structure around dislocations, their presence in multiphase materials causes heterogeneous nucleation of the product phase and thus also shifts of the transformation temperature. This novel field-theoretical approach is very convenient as it allows to bridge the gap in studying the behavior of materials at the length and time scales that are not attainable by atomistic or macroscopic models.

## Introduction

Physical properties of solids are invariably affected by crystal defects and their associated long-range strain fields that distort the underlying crystal lattice. The lattice thus represents an elastic template whose distortions affect those physical properties of solids that are sensitive to the crystal symmetry. This is typical for ferroelastic martensites, where lattice distortions due to defects shift the transformation temperature and thus allow for heterogeneous nucleation of the low-symmetry product phase in the matrix of the parent (typically cubic) phase. Similarly, in piezoelectric materials the polarization couples to strain and, therefore, any distortion of the lattice and thus the onset of spontaneous strain necessarily lead to changes in the polarizability of the material. To guarantee the reliability of these devices, it is necessary to understand how coupling between microstructure and defects affect the physical properties of interest. This opens a possibility of a systematic strain engineering by which new materials emerge with desired physical properties.

Coupling of the microstructure with plasticity can be accomplished efficiently using the mesoscopic model of dislocation-mediated phase transformations that was published recently in [1]. The structural part of this model is described using the Landau-Ginzburg theory of the first order phase transitions that has been shown to be valid across many length scales down to a few lattice spacings (for a reference, see [2]). The dislocation network is incorporated utilizing Kröner's continuum theory of dislocations [3]. Within this theory, dislocations are viewed as sources of incompatible plastic deformation that requires elastic relaxation to keep the body compact. Our objective in this paper is to investigate how a single edge dislocation in a periodic body affects the spatial variation of the microstructure in materials that undergo a cubic to tetragonal (orthorhombic) phase transformation. These calculations are carried out both above and below the transformation temperature $T_c$ to reveal the characteristic distortions around the dislocation in spatially uniform (austenitic) and twinned (martensitic) microstructures.

**The free energy functional**

In the following, we will be concerned with a two-dimensional representation of a cubic to tetragonal (or orthorhombic) phase transformation. This is driven by a soft phonon mode whose frequency becomes imaginary as the temperature drops below the transformation temperature $T_c$. The change of the free energy density for small elastic distortions of the parent cubic phase is equal to the elastic strain energy density, i.e. $\Delta G = (1/2)c_{ijkl}\varepsilon_{ij}\varepsilon_{kl}$, where $c_{ijkl}$ is the elastic stiffness tensor and $\varepsilon_{ij}$ the components of the elastic strain tensor. This free energy can be further simplified by defining a set of symmetry-adapted strain order parameters $e_i$ that represent the three fundamental modes of deformation [4]. These are typically written as $e_1 = (\varepsilon_{11} + \varepsilon_{22})/\sqrt{2}$, $e_2 = (\varepsilon_{11} - \varepsilon_{22})/\sqrt{2}$, and $e_3 = \varepsilon_{12}$. Here, $e_1$ represents a hydrostatic deformation, $e_2$ the change of phase from the parent cubic structure to the product tetragonal (or orthorhombic) structure ($e_2 \neq 0$), and $e_3$ the shear of the crystal lattice. Since we are concerned with the change of phase, our primary order parameter is $e_2$, while $e_1$ and $e_3$ are secondary order parameters. Replacing the strains in the free energy by these order parameters yields the harmonic form

$$\Delta G = \frac{A_1(T)}{2} e_1^2 + \frac{A_2(T)}{2} e_2^2 + \frac{A_3(T)}{2} e_3^2, \tag{1}$$

where the coefficients $A_1 = C_{11} + C_{12}$, $A_2 = C_{11} - C_{12}$ and $A_3 = 4C_{44}$ are functions of the elastic constants and, therefore, they generally depend on temperature and, for large pressures, also on pressure. The minimum of $\Delta G$ corresponds to the parent cubic phase and so the form (1) does not allow for any phase change. These phase changes can be modeled by regarding the free energy as an expansion in terms of the primary order parameter. Only the terms that are independent with respect to all generators of the point group of the parent cubic lattice are retained in this expansion. In our case, it means that the free energy can contain only even powers of the primary order parameter $e_2$. If this expansion is terminated by the fourth-order term, the resulting free energy functional will model the second order (continuous) phase transition. On the other hand, adding both the fourth and the sixth order terms yields the free energy specific to the first-order (abrupt) phase transition that gives rise to deformation twinning in ferroelastic martensites. For convenience, we split the free energy density into a part that depends explicitly on the primary order parameter, $f_{loc} = [A_2(T)/2]e_2^2 + (B/4)e_2^4 + (C/6)e_2^6$, and a term that contains only the secondary order parameters, $f_{nonloc} = [A_1(T)/2]e_1^2 + [A_3(T)/2]e_3^2$. We will show below that the latter term gives rise to nonlocal interactions in the microstructure. Finally, we incorporate the length scale by including the Ginzburg gradient term of the primary order parameter, $f_{grad} = (K_2/2)(\nabla e_2)^2$. The total free energy density then reads $f = f_{loc} + f_{nonloc} + f_{grad}$.

In the following, we will consider that the crystal contains an arbitrary but known dislocation network. In order to ensure that no microcracks form during the minimization of the free energy and the buildup of spontaneous strain, the components of the internal strain field have to satisfy the incompatibility condition [3]. Mathematically, it means that $\nabla \times \nabla \times \boldsymbol{\varepsilon} = \boldsymbol{\eta}$ has to be satisfied at each point of the body. Here, $\boldsymbol{\eta} = \mathrm{sym}\,(\nabla \times \boldsymbol{\alpha})$ is the incompatibility tensor and $\boldsymbol{\alpha}$ the Nye tensor that represents the density of continuous distribution of infinitesimal Burgers vectors. In general, this condition represents six coupled (but not independent) partial differential equations. However, in two-dimensional plane strain problems only one equation is not satisfied trivially and this represents the incompatibility condition that constrains the spatial variation of the internal strain field. The reason is that plane strain problems in the $(x_1, x_2)$ plane can only accommodate edge dislocations with their line directions parallel to $x_3$. Consequently, the two nonzero components of the Nye tensor are $\alpha_{31}$ and $\alpha_{32}$ and thus the only incompatibility equation that is not satisfied trivially is that for $\eta_{33}$. This can be written as [5] $\partial_{22}\varepsilon_{11} - 2\partial_{12}\varepsilon_{12} + \partial_{11}\varepsilon_{22} = \eta_{33}$, where $\eta_{33} = \partial_1\alpha_{32} - \partial_2\alpha_{31}$. If this constraint is expressed in the form $g = 0$, it can be incorporated into the free energy functional using a Lagrange multiplier $\lambda$. The free energy is thus:

$$F = \int_S (f_{loc} + f_{nonloc} + f_{grad} + \lambda g)\, \mathrm{d}S \quad . \tag{2}$$

Since in the stationary state $\delta F/\delta e_1 = \delta F/\delta e_3 = \delta F/\delta \lambda = 0$, one can eliminate from the free energy the secondary order parameters (for details, see [1]) which gives rise to nonlocal interactions in the microstructure. The free energy is thus a functional of the primary order parameter $e_2$ only. Hence, the evolution of $e_2$ (and thus also of the microstructure) towards the equilibrium is obtained by solving the Landau-Khalatnikov equation $\partial e_2/\partial t = -\gamma \delta F/\delta e_2$, where $\gamma$ is a mobility parameter.

**Effect of an edge dislocation on the spatial variation of the microstructure**

We will now utilize the free energy (2) to calculate the equilibrium spatial variation of the microstructure above $T_c$ (austenite) and below $T_c$ (martensite). The two-dimensional simulated block is oriented such that the axis $x_1$ is parallel to the horizontal [10] direction, $x_2$ is parallel to the vertical [01] direction, and $x_3$ is perpendicular to both $x_1$ and $x_2$. The parameters of the free energy correspond to Fe-30at.%Pd and are the same as in [1]. For simplicity, we will focus on two cases: (i) defect-free medium, and (ii) material containing one edge dislocation with its Burgers vector parallel to $x_1$ and line direction parallel to $x_3$ inserted in the middle of the simulated cell. The equilibrium fields $e_2$ for these two cases obtained by minimizing the free energy above $T_c$ are shown in Fig. 1 and below $T_c$ in Fig. 2. In both figures, middle gray corresponds to $e_2 = 0$ (cubic austenitic phase), lighter grays to white to $e_2 > 0$ and darker grays to black to $e_2 < 0$ (the two martensite variants). The range of shadows of gray is identical in all these figures.

For a dislocation-free medium above $T_c$, one obtains a uniform microstructure of the austenite (Fig. 1a). Inserting an edge dislocation into austenite produces lattice distortions of the usual butterfly shape which is shown in Fig. 1b. Since the dislocation is created by sliding into the block a half-plane $(x_2, x_3)$ from the top, the part of the lattice at $x_2 > 0$ close to the dislocation is compressed in the $x_1$ direction (dark gray). Similarly, the dislocation makes the lattice at $x_2 < 0$ expand in the $x_1$ direction, which is shown in Fig. 1b by lighter shades of gray. The lattice distortions induced by the dislocation thus represent sites for heterogeneous nucleation of the low-symmetry phase. This allows for shifts of $T_c$ above its value for the dislocation-free medium.

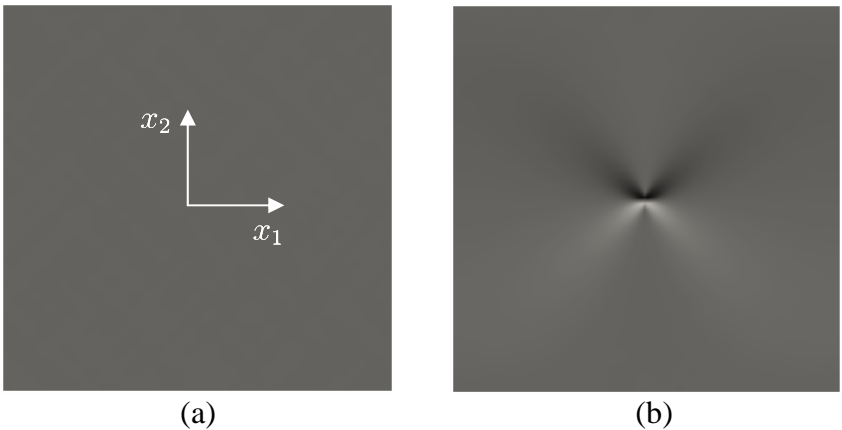

Fig. 1: Spatial variation of the order parameter field $e_2$ *above $T_c$* for a body without dislocation (a) and with an edge dislocation in the middle (b).

(a)    (b)

Below $T_c$ the system undergoes spontaneous twinning shown in Fig. 2a with the normals of the twins parallel to $\langle 11 \rangle$ directions and employing both variants of the low-symmetry phase (light gray and black). The parent cubic phase is now metastable and the twin boundaries are decorated by the austenite separating the two different variants of the martensite. For a dislocation-free material, one obtains the typical twinned microstructure shown in Fig. 2a. Due to the symmetry of the block, this pattern is degenerate under the group operations of the square. When the dislocation is inserted into the middle of the block, the twinned microstructure is affected by the strain field of the dislocation. This is shown in Fig. 2b, where the twin boundary passess exactly through the lattice site into which the dislocation was inserted. To minimize the energy cost $f_{grad}$, the system accommodates the dislocation strain field by merging it as much as possible with that due to the twins. This is possible at points marked $A+$ and $A-$, where the presence of the dislocation results in local bending of the twin boundary. On the contrary, a frustration between the strain field of the dislocation and the twinned microstructure results in strain-induced switching of the two martensite variants at $B+$ and $B-$. Minimization of the cost of embedding the two lobes in the region of a different martensite variant results in bending of the perpendicular twin boundaries in the vicinity of the tips of these lobes ($C$). Finally, one can see that the surplus energy due to a locally disordered field $e_2$ in the middle of Fig. 2b is compensated by reducing the number of twin boundaries and thus coarsening the twinned microstructure.

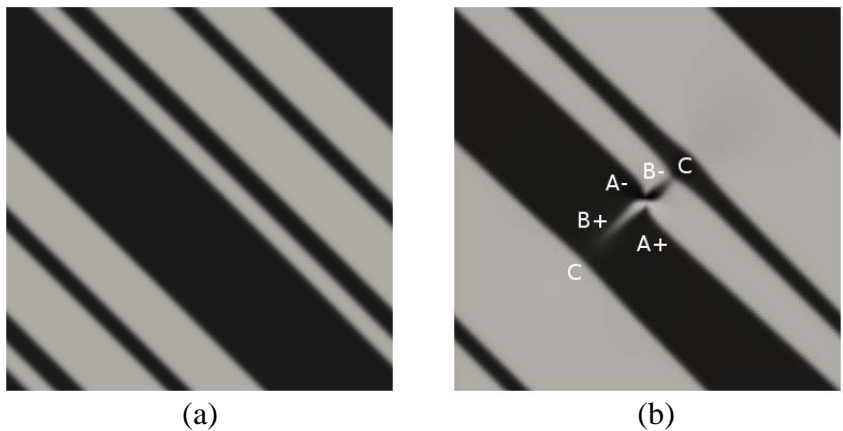

(a)   (b)

Fig. 2: Spatial variation of the order parameter field $e_2$ *below $T_c$* for a body without dislocation (a) and with an edge dislocation in the middle (b).

**Summary**

We have shown how Kröner's concept of the incompatibility of strains can be coupled with the strain-based Landau-Ginzburg free energy functional. This gives rise to a mesoscopic model of ferroelastic materials in which the spatial variation of the internal strain field is constrained by the preexisting dislocation network. The fields of dislocations are capable of large changes in the microstructure that may deteriorate the mechanical properties of ferroelastic materials. In piezoelectric and piezomagnetic crystals, the coupling of the strain field to polarization or magnetization will cause local depolarization or demagnetization of the lattice.

**Acknowledgements**

This research was supported by the Marie-Curie International Reintegration Grant No. 247705 "MesoPhysDef", and in part by the Academy of Sciences of the Czech Republic (Research Project No. AV0Z20410507) and by the U.S. Department of Energy.

**References**

[1]   R. Gröger, T. Lookman, and A. Saxena: Phys. Rev. B 78 (2008), no. 184101.
[2]   E. K. H. Salje: *Phase transitions in ferroelastic and co-elastic crystals* (Cambridge University Press, 1993).
[3]   E. Kröner: *Continuum theory of dislocations and self-stresses* (Springer-Verlag, 1958).
[4]   G. R. Barsch and J.A. Krumhansl: Phys. Rev. Lett. 53 (1984), p. 1069.
[5]   T. Lookman, S. R. Shenoy, K. Ø. Rasmussen, A. Saxena, and A. R. Bishop: Phys. Rev. B 67 (2003), no. 024114.